\documentclass{llncs}
\usepackage{makeidx}  % allows for indexgeneration
\usepackage{graphicx}
\usepackage[]{graphicx}
\usepackage{amsmath }
\usepackage{color}
\begin{document}
\mainmatter              % start of the contributions

\title{
Improving Surgical Training  Phantoms by Hyperrealism: Deep Unpaired Image-to-Image Translation from Real Surgeries}

\titlerunning{Hyperrealistic Surgical Training}  
\author{Sandy~Engelhardt\inst{1,3} \and
Raffaele~De~Simone\inst{2} \and
Peter M. Full\inst{2} \and
Matthias~Karck\inst{2}\and
Ivo~Wolf\inst{3}}

\authorrunning{Engelhardt et al.} % abbreviated author list (for running head)
%\tocauthor{Sandy Engelhardt, Raffaele De Simone, Peter M. Full, Matthias Karck, and Ivo Wolf}
%
\institute{Dep. of Simulation and Graphics, Magdeburg University, Germany\\
\email{sandy.engelhardt@isg.cs.uni-madgeburg.de}
\and
Department of Cardiac Surgery, Heidelberg University Hospital, Germany
\and
Faculty of Computer Science, Mannheim University of Applied Sciences, Germany}

\maketitle              % typeset the title of the contribution

\begin{abstract} %(2000 char)
Current `dry lab' surgical phantom simulators are a valuable tool for surgeons which allows them to improve their dexterity and skill with surgical instruments. These phantoms mimic the haptic and shape of organs of interest, but lack a realistic visual appearance.
In this work, we present an innovative application in which representations learned from real intraoperative endoscopic sequences are transferred to a surgical phantom scenario. 
The term \textit{hyperrealism} is introduced in this field, which we regard as a novel subform of surgical augmented reality for approaches that involve real-time object transfigurations. 
For related tasks in the computer vision community, unpaired cycle-consistent Generative Adversarial Networks (GANs) have shown excellent results on still RGB images. Though, application of this approach to continuous video frames can result in flickering, which turned out to be especially prominent for this application. Therefore, we propose an extension of cycle-consistent GANs, named \textit{tempCycleGAN}, to improve temporal consistency.
The novel method is evaluated on captures of a silicone phantom for training endoscopic reconstructive mitral valve procedures.
Synthesized videos show highly realistic results with regard to 1) replacement of the silicone appearance of the phantom valve by intraoperative tissue texture, while 2) explicitly keeping crucial features in the scene, such as instruments, sutures and prostheses. Compared to the original CycleGAN approach, \textit{tempCycleGAN} efficiently removes flickering between frames.
The overall approach is expected to change the future design of surgical training simulators since the generated sequences clearly demonstrate the feasibility to enable a considerably more realistic training experience for minimally-invasive procedures.

\end{abstract}

\section{Introduction}
\begin{figure}[t]
   \begin{center}
 \includegraphics[width=\linewidth]{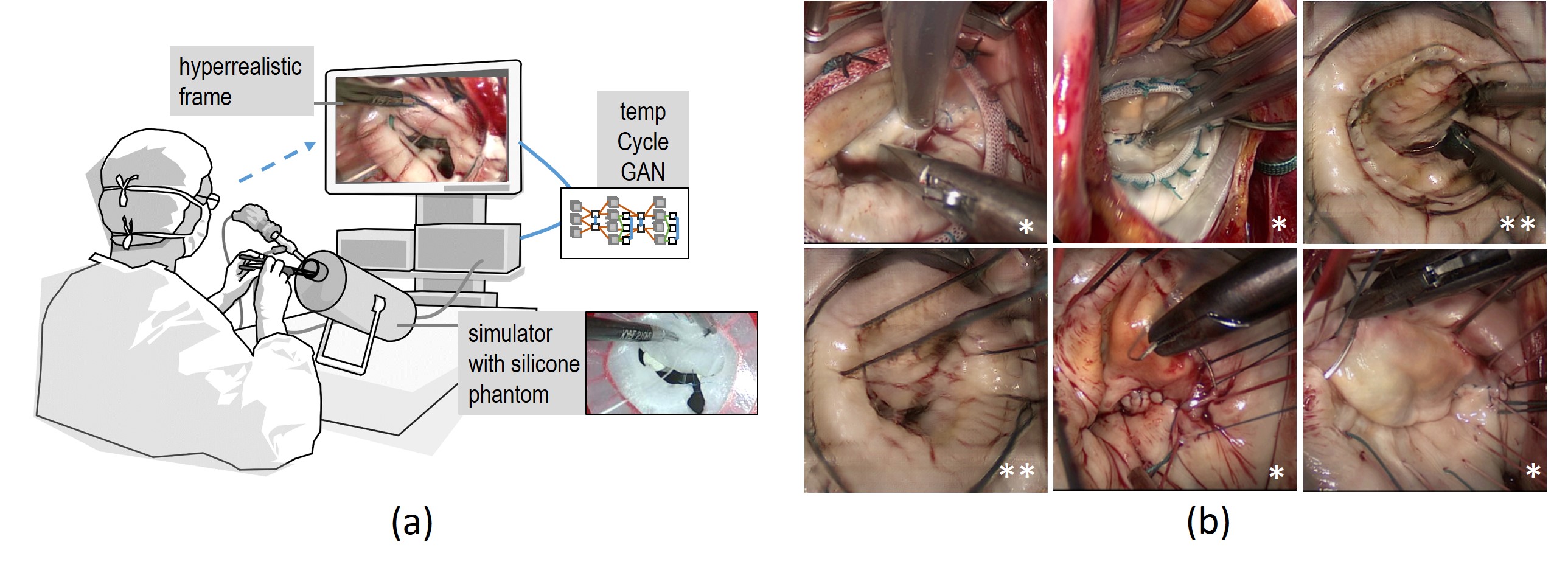}
   \end{center}
   \caption[] 
   { \label{fig:Illu} (a) Vision: Augmentation of the minimally invasive training process with real-time generated \textit{hyperrealistic} frames. (b) Visual comparison of real intraoperative frames from mitral valve surgery (*) and generated fake images (**).}
\end{figure}

Surgery is a discipline that requires years of training to gain the necessary experience, skill and dexterity. With increasingly minimally invasive procedures, in which the surgeon's vision often solely relies on endoscopy, this is even more challenging. 
Due to the lack of appropriately realistic and elaborate endoscopic training methods, surgeons are forced to develop most of their skills in patients, which is truly undesirable.
Current training methods rely on practising suturing techniques on \textit{ex-vivo} organs (`wet labs'), virtual simulators or physical phantoms under laboratory conditions (`dry labs'). Training on authentic tissue is associated with organizational efforts and costs and is usually not accessible to the majority of the trainees. Virtual simulators overcome these requirements, but are often less realistic due to the lack of blood, smoke, lens contamination and patient-specificity. Physical phantoms, e.g. made from silicone, suffer also from these drawbacks, but they provide excellent haptic feedback and tissue properties for stitching with authentic instruments and suture material \cite{Kenngott2015,Engelhardt2018}.
However, their uniform appearance does not reflect the complex environment of a surgical scene. We tackle this issue by proposing a system that is able to map patterns learned from intraoperative video sequences onto the video stream captured during training with silicone models to mimic the intraoperative domain. Our vision for a novel training simulator is to display real-time synthesized images to the trainee surgeon while he/she is operating on a phantom under restricted direct vision, such as illustrated in Fig. \ref{fig:Illu}a.

Generative Adversarial Networks (GANs) demonstrate tremendous progress in the field of image-to-image translation with regard to both perceptual realism and diversity. 
Recently, methods have been proposed using Convolutional Neural Networks (CNNs) for deep image synthesis with paired \cite{Isola2017} and even unpaired natural images, namely DualGAN \cite{Yi2017} and CycleGAN \cite{Zhu2017}. These networks translate an image from one domain X to another target domain Y. The key to the success of GANs is the idea of an adversarial loss that forces the generated images to be, in principle, indistinguishable from real images, which is particularly powerful for image generation tasks. However, current solutions do not take time consistency of a video stream into account. While each frame of a generated video looks quite realistic on its own, the whole sequence lacks consistency. 

In order to increase realism for endoscopic surgical training on physical phantoms, we propose the concept of \textit{hyperrealism}.\footnote{The term is related to the homonymous art form, where an excessive use of details is used to create an exaggeration of reality which cannot be seen by the human eye.} 
We define hyperrealism as a new paradigm on the Reality-Virtuality continuum \cite{milgram_taxonomy_1994} as a concept closer to `full reality' in comparison to other applications where artificial overlays are superimposed on a video frame. 
In a hyperrealistic environment, those parts of the physical phantom that look unnatural are replaced by realistic appearances.

The extended CycleGAN network, named \textit{tempCycleGAN}, learns to translate an image stream from the source domain of \textit{phantom data} to a target domain \textit{intraoperative surgeries} and vice versa in the absence of paired endoscopic examples.
The network's main task is to capture specific characteristics of one image set and to figure out how these characteristics could be translated into the other image domain.
We evaluate the approach for the specific application of training mitral valve repair, where the network has to learn 1) how to enhance the silicon's surface appearance, at the same time not altering its shape,
2) not to replace other important features in the scene, such as surgical instruments, sutures, needles and prostheses.

\section{Methods}

\begin{figure}[t]
   \begin{center}
 \includegraphics[width=0.8 \linewidth]{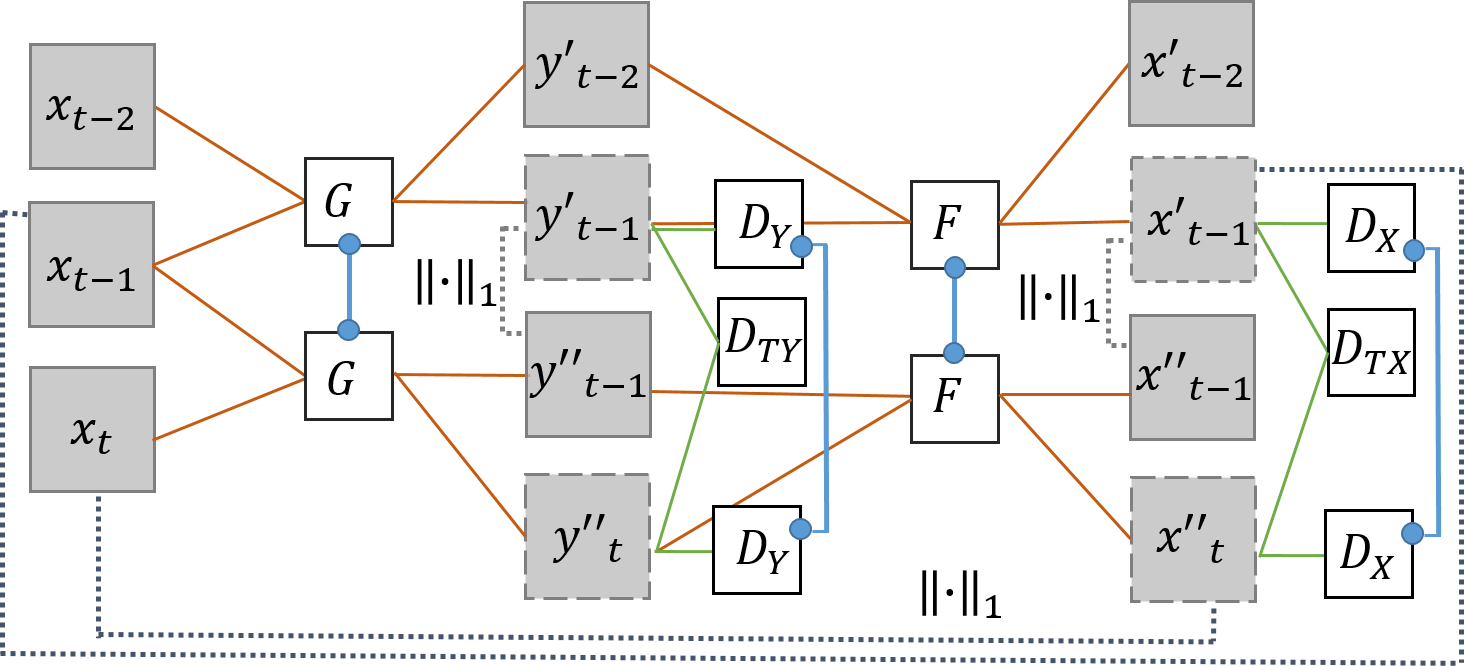}
   \end{center}
   \caption[] 
   { \label{fig:Archi} 
   Training setup of the $X \to Y \to X$ cycle of the proposed tempCycleGAN network (reverse cycle accordingly) using temporal pairs: the generators $G$, $F$ and the temporal discriminators $D_{T\{X,Y\}}$ take the current frame and a single preceding frame.
   Each run of $G$ (and $F$) synthesizes outputs for both frames.
   In the application of the generator, the frame of interest is the second output ($y'_{t-1}$ and $y''_{t}$, respectively).
   The temporal discriminators are trained on these frames of interest, 
   thus, the generator $G$ needs to run twice to generate the two frames of interest ($y'_{t-1}$ and $y''_{t}$) for $D_{TY}$ (for $F$ and $D_{TX}$ accordingly).
   L1-distances (dotted lines) between matching time frames are used in the loss function to further enforce time consistency. 
   $D_{\{X,Y\}}$: discriminators with 1 input; blue connections: shared weights.}
\end{figure} 

We build upon the CycleGAN model proposed by Zhu et al. \cite{Zhu2017}. % (concurrently developed as DualGANs in \cite{Yi2017}).
The goal of CycleGANs is to obtain mapping functions between two domains X and Y given unpaired training samples, $\{x_i|i=1..N\}\in X$ and $\{y_j|j=1..M\}\in Y$.
Mappings in both directions are learned by two generator networks, $G: X\to Y$ and $F: Y\to X$.
The generators are trained to produce output that cannot be distinguished from real images of the target domain by adversarially trained discriminators $D_Y$ and $D_X$.

\subsection{Temporal Cycle GAN}
Temporal consistency requires to include preceding time steps into the learning process. 
The proposed advancement tempCycleGAN processes the current time step $x_t$ and its two predecessors $x_{t-1}$ and $x_{t-2}$, as shown in Fig.~\ref{fig:Archi}. 
In general, it is possible to use more preceding frames and to adjust the network architecture accordingly.
In the following, the general concepts of tempCycleGAN are explained and a detailed description of the setup is provided  subsequently. 

Temporal discriminators $D_{TX}$ and $D_{TY}$ (one for each domain) are introduced that take consecutive frames and try, as usual, to distinguish real from generated data.
The idea is that flickering would allow the discriminators to easily identify generated data.
Thus, the generators are forced to avoid flickering to successfully cheat their adversarial temporal discriminators.
The generators need at least one preceding frame as additional input to be able to create a temporal consistent output for the current frame.
In the current setup two frames are used as input for both the generators and the temporal discriminators.

To define a cycle consistency loss that is symmetric in $G$ and $F$, we let the generators $G$ and $F$ create as many output frames as they get input frames. 
For example, $G(x_{t-1}, x_t)$ creates outputs $y''_{t-1}$ and $y''_t$. 
Only the output for the latest frame ($y''_t$ in the example) is the frame of interest used in the actual output video. 
Consecutive frames of interest (shown as dashed boxes in Fig.~\ref{fig:Archi}) are evaluated by the temporal discriminators.
Thus, the temporal discriminators are provided with inputs of multiple runs of the generator.
For example, $D_{TY}$ takes $y'_{t-1}$ and $y''_t$ as input, where $y'_{t-1}$ is the frame of interest of $G(x_{t-2}, x_{t-1})$ and $y''_{t}$ is the frame of interest of $G(x_{t-1}, x_{t})$.
To enforce consistency between frames of matching time and domain, L1 distances to the respective frames are used as additional terms in the loss function.

\subsection{Network Architectures}
The network architectures of the generators and discriminators are largely the same as in the original CycleGAN approach \cite{Zhu2017}.
A TensorFlow implementation provided on GitHub\footnote{https://github.com/LynnHo/CycleGAN-Tensorflow-PyTorch-Simple}
was used as the basis and extended with the new tempCycleGAN blocks.
All discriminators take the complete input images, which is different from the $70 \times 70$ PatchGAN approach by Zhu et al. \cite{Zhu2017}.
The temporal discriminators have 6 ($2\times$RGB) instead of 3 input channels.
For the generators, 8 instead of 9 residual blocks are used, because experiments on our data showed better results for this configuration.

\section{Experiments}

The commercial minimally invasive mitral valve simulator (MICS MVR surgical simulator, Fehling Instruments GmbH \& Co. KG, Karlstein, Germany) was extended with patient-specific silicone valves. Details on 3D-printed mold and valve production are elaborated on in a previous work \cite{Engelhardt2018}.
An expert segmented mitral valves with different pathologies, such as posterior prolaps and ischemic valves on the end-systolic time step from echocardiographic data. From these virtual models, 3D printable molds and suitable annuloplasty rings were automatically generated with stitching holes using 3 different low to medium cost 3D-printers, varying material (polylactide, acrylonitrile butadiene styrene) in various colors (e.g. white, beige, red, orange). From these molds, 10 silicone valves were cast that could be anchored in the simulator on a printed valve holder. We asked an expert and a trainee to apply mitral valve repair techniques (annuloplasty, triangular leaflet resection, neo-chordae implantation) on these valves and captured the training process endoscopically.   

\subsection{Data and Training of Network}
In total, approx.\ 330,000 video frames from the training procedures in full HD resolution were captured.
Valves shown in videos for training were not used for testing. 
For training, three continuous frames after each 120th frame from a subset of approx.\ 160,000 frames was sampled retrospectively, such that the set comprised 1300 small sequences.
Furthermore, training material for the target domain from 3 endoscopic mitral valve repair surgeries was captured. 
In total, approx.\ 320,000 frames were acquired during real surgery. For training, three continuous frames after each 240th frame were sampled retrospectively and 1294 small sequences were used for training. All streams were captured with 30fps. The scenes are highly diverse, as the valve's appearance drastically changes over time (e.g. due to cutting of tissue, implanting sutures and prostheses, fluids such as blood and saline solution), see Fig. \ref{fig:Illu}b. 
All frames were square-cropped and re-scaled to 286 $\times$ 286. 
Data augmentation was performed by random cropping of a $256 \times 256$ region and random horizontal flipping.  
For all the experiments, the consistency loss was weighted with $\lambda = 10$ \cite{Zhu2017}. The Adam solver with a batch size of 1 and a learning rate of 0.0001 without linear decay to zero was used. Similar to Zhu et al. \cite{Zhu2017}, the objective was divided by 2 while optimizing $D$, which slows down the rate at which $D$ learns relative to $G$. Discriminators are updated using a history of 50 generated images rather than the ones produced by the latest generative networks \cite{Zhu2017}. The tempCycleGAN network was trained for 40, 60, 80, 100 epochs to find the visually most attractive results. In analogy, the original CycleGAN networks was trained either with 1 input frame or 3 continuous frames.  

\subsection{Evaluation}

The most important factors for the proposed application are related to perception i.e. how real the generated intraoperative videos appear to an expert with years of experience in mitral valve surgery. Furthermore, reliability plays a crucial role, as the appearance of the scene should be transferred into the target domain, while neither the shape of objects should be altered, nor additional parts should be added or taken away.   

\textit{Realness:} An expert was asked to score the visual quality of eleven 10s mini videos synthesized from the test set phantom frames. For assessment, the ``realness score" was used, as proposed by Yi et al. \cite{Yi2017}, ranging from 0 (totally missing), 1 (bad), 2 (acceptable), 3 (good), to 4 (compelling). We decided against the conduction of a Visual Turing Test, as some shape-related features in the scene (a personalized ring shape instead of a standard commercial ring was used in the experiments) would have been easily identified by an expert surgeon. 

\textit{Reliability:} The result of tempCycleGAN was comprehensively compared in terms of faithfulness to the input phantom images using 39 randomly selected synthesized frames from the test set (16 showing an annuloplasty and 23 showing a triangular leaflet resection). Predefined criteria relevant for surgery were assessed, e.g. whether the instruments or all green and white sutures are completely visible or whether artifacts disturb the surgical region of interest. 
12 of 39 input frames show two instruments and three frames only a single instrument. Stitching needles held by the instrument are used in six frames and a prosthetic ring is visible in 29 frames. On average, $4.9$ white and $4.9$ green sutures are observable in the phantom frames.

\section{Results}

\begin{figure}[t]
   \begin{center}
 \includegraphics[width=\linewidth]{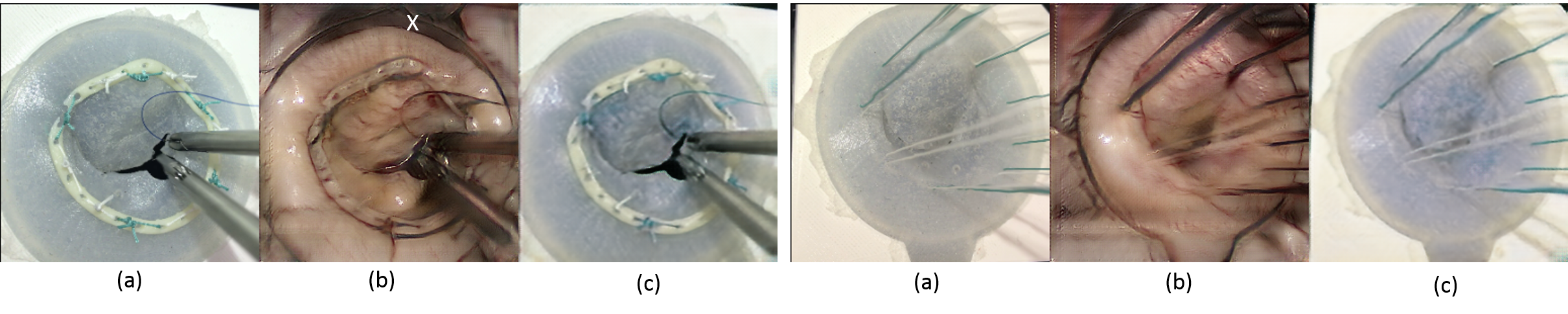}
   \end{center}
   \caption[] 
   { \label{fig:Result} tempCycleGAN results for two examples shown left and right, where (a) shows the real phantom $x_t$, (b) shows corresponding synthesized intraoperative images $y''_t$ and (c) shows the re-synthesized phantom image $x''_t$. X marks a synthesized atrial retractor.}
\end{figure}

The result of tempCycleGAN was visually most attractive after 60 epochs. Examples are provided in Fig. \ref{fig:Result}.
Model training of the tempCycleGAN took 18 hours for 60 epochs on a single NVIDIA GeForce Titan Xp GPU.
Compared to the original CycleGAN \cite{Zhu2017} trained with a single frame or three consecutive frames, tempCycleGAN produces results with no flickering, contains fewer artifacts, and better preserves content structures in the inputs and capture features (e.g., texture and/or color) of the target domain (see supplemental material videos).
The tempCycleGAN approach was even capable of learning where semantically to insert blood (between the leaflets) or an atrial retractor in the scene (Fig. \ref{fig:Result}). However, it produced slightly blurred instruments and sutures.   

The average ``realness score" of the 11 mini-videos assessed by the expert surgeon was 3.3 (5$\times$ category  ``compelling", 4 $\times$  ``good", 2 $\times$ ``acceptable"). Longer versions of a compelling scene (first scene) and an acceptable scene (second scene) are provided in the supplemental video showing a ring annuloplasty\footnote{https://youtu.be/qugAYpK-Z4M}.  
The valve's texture was assessed as very realistic in general by the surgeon. Some instruments and rings appeared blurry and had minor artifacts (e.g. the projection of the sewing cuff of the original ring onto the printed ring appeared incomplete), which led to a lower realness score. However, for most of the scenes this was not crucial because relevant image regions were not effected. 

The quantitative assessment of reliability (i.e. comparison of source and target frame) yielded different results for instruments, needles, sutures and silicone surface: Neither instruments, needles nor annuloplasty rings were erroneously added. One generated instrument was classified as `not preserved', since it was partially coalesced with the valve and two (out of six) needles could not be seen in the generated images. Green sutures were better preserved ($4.0$ of $4.9$ sutures per frame) compared to white sutures ($2.2$ of $4.9$ sutures per frame). In 14 frames all green sutures were consistent in both source and target domain, whereas there is no such frame for white sutures.  
The appearance of the generated frames was evaluated to be `overall realistic' in $82.1\%$. The quality of the generated valves (shape and tissue texture) was compared to the silicone valve. The visual inspection yielded `valve differs completely' in $2.6\%$, `good alignment but details differ' in $10.3\%$ and `good agreement' in $87.2\%$. 

\section{Discussion}
According to the widely accepted definition from Azuma \cite{Azuma1997}, Augmented Reality (AR) ``\textit{allows the user to see the real world, with virtual objects superimposed upon or composited with the real world.  Therefore, AR supplements reality, rather than completely replacing it. Ideally, it would appear to the user that the virtual and real objects coexisted in the same space"}.
We consider \textit{hyperrealism} as a subform of AR where real, but artificially looking objects (in our case the silicone valves) are transfigured to appear realistically (as in a real surgery). Nothing is added to the scene of the real world, it is just altered to appear more realistic, thus the term hyperrealistic. Objects that already appear realistic ideally stay the same (in our case the instruments, sutures, needles).

The idea to use a transformer network to translate a real endoscopic image into a synthetic-like virtual image has been assessed before with the overall aim of obtaining a reconstructed topography \cite{Visentini-Scarzanella2017,Mahmood2017}. We focus on the opposite transformation, synthesizing intraoperative images from real training procedures on patient-specific silicone models. Our scenes are more complex, since they contain e.g. blood and lens contamination in the target domain and moving instruments, sutures and needles in both source and target domains.

Our methodological advancement tempCycleGAN shows a substantially stabilized composition of the synthesized frames in comparison to the original CycleGAN approach.
The architecture's extension by two temporal discriminators, temporal paired input frames fed into multiple runs of generators and further L1 distances in the loss function to penetrate inconsistency yields such significantly more stable results. Beyond that, tempCycleGAN reduced the number of artifacts in the reported outcomes, while slightly sacrificing image sharpness. 
To the best of our knowledge, our approach is the first method for unpaired image-to-image translation addressing the problem of temporal inconsistencies in moving sequences.     

\section*{Acknowledgements}
The authors thank Bernhard Preim for his valuable hints and Benjamin Hatscher for making the illustration in Fig. \ref{fig:Illu}a.
The work was supported by DFG grant DE 2131/2-1, EN 1197/2-1.
The GPU was donated by NVidia small scale grant.

\end{document}